\documentclass[twocolumn]{aastex62}
\usepackage{graphicx}
\usepackage{amssymb}                    
\usepackage{ulem}
\usepackage{float}
\usepackage{rotating}
\usepackage{enumitem}
\usepackage{rotating}
\usepackage{makecell}
\usepackage{appendix}
\usepackage{fixmath}
\usepackage{epstopdf}
\usepackage{times} 
\usepackage{amsmath}
\usepackage{booktabs}

\definecolor{green1}{rgb}{0.2, 0.6, 0.1}
\definecolor{purp1}{rgb}{0.6, 0.1, 0.45}

\newcommand{\hic}{Hi-C\,2.1}
\newcommand{\ar}{AR\,12712}

\newcommand{\arc}{\ensuremath{^{\prime\prime}}}

\newcommand{\cf}{non-linear least-squares curve fitting method}

\received{6 April 2020}
\submitjournal{ApJ}

\shortauthors{Williams et al. (2020)}
\shorttitle{Analysis of Hidden Strand Widths}
\begin{document}

\title{Evidence for and Analysis of Multiple Hidden Coronal Strands in Cross-Sectional Emission Profiles: Further Results from NASA's High-resolution Solar Coronal Imager}
\author{Thomas Williams}
\affil{Jeremiah Horrocks Institute, University of Central Lancashire, Preston, PR1 2HE, UK}
\author{Robert W.\ Walsh}
\affil{Jeremiah Horrocks Institute, University of Central Lancashire, Preston, PR1 2HE, UK}
\author{Hardi Peter}
\affil{Max Planck Institute for Solar System Research, Justus-von-Liebig-Weg 3, D-37077, G\"{o}ttingen, Germany}
\author{Amy R.\ Winebarger}
\affil{NASA Marshall Space Flight Center, ST13, Huntsville, AL 35812, USA}

\begin{abstract}
Previous work utilising NASA's High-resolution Coronal Imager (\hic) 172\,\AA\ observations revealed that, even at the increased spatial scales available in the data-set, there may be evidence for coronal structures that are still not fully resolved. In this follow-up study, cross-section slices of coronal strands are taken across the \hic\ field-of-view. Following previous loop width studies, the background emission is removed to isolate the coronal strands. The resulting intensity variations are reproduced by simultaneously fitting multiple Gaussian profiles using a non-linear least-squares curve fitting method. In total, 183 Gaussian profiles are examined for possible structures that are hinted at in the data. The full width at half maximum (FWHM) is determined for each Gaussian, which are then collated and analysed. The most frequent structural widths are $\approx450-575$\,km with 47\% of the strand widths beneath NASA's Solar Dynamics Observatory Atmospheric Imaging Assembly (AIA) resolving scale (600-1000 km). Only 17\% reside beneath an AIA pixel width (435\,km) with just 6\% of the strands at the \hic\ resolving scale ($\approx220-340$\,km). These results suggest that non-Gaussian shaped cross-sectional emission profiles observed by \hic\ are the result of multiple strands along the integrated line-of-sight that can be resolved, rather than being the result of even finer sub-resolution elements.
\end{abstract}

\section{Introduction}
Observational investigations of coronal loop structure have been undertaken since the 1940s \citep{bray91}; however, due to insufficient spatial resolution of current and previous instrumentation, the definitive resolved widths of these fundamental structures have not been fully realised. Recent high-resolution data from NASA's Interface Region Imaging Spectrometer (IRIS; \citealp{depontieu14}) and the High-resolution Coronal imager (Hi-C; \citealp{kobayashi14}) have led to coronal loop width studies in unprecedented detail. For short loops whose lengths are of the scale of a granule, \citet{peter13} found widths $\lesssim200$\,km within the Hi-C data. Similarly, \citet{aschwanden17} sampled $10^5$ loop width measurements from the Hi-C field of view (FOV) with their analysis finding the most-likely width $\approx550$~km, arguing the possibility that Hi-C fully resolved the 193\,\AA\ loops/strands. This agrees with previous work \citep{peter13} where it is proposed that at least some of the wider loops with diameters $\approx$1\,Mm observed by NASA's Solar Dynamic Observatory Atmospheric Imaging Assembly (AIA; \citealp{lemen12}) do not appear to show what they consider to be obvious signs of substructure when compared to the coincident Hi-C data-set. Combining IRIS data with hydrodynamic simulations, \citet{brooks16} find transition region temperature loops with widths between $266-386$\,km, and showcase that these structures appear to be composed of singular magnetic threads.

\citet{klimchuk15} investigates the widths of four EUV loops as a function of position using Hi-C and AIA data, and obtain widths of $880-1410$\,km with Hi-C. They also find that, while the analysed loops have relatively constant cross-section along their lengths, those measured with Hi-C  are typically less than 25\% narrower than their AIA counterparts. Therefore, they suggest that loops are not highly under-resolved by AIA and these results further supports previous findings of measured widths along both EUV \citep{fuentes06} and soft x-ray \citep{klimchuk92,klimchuk00} loop structures where no significant or observable expansion from the loop base to apex are determined.This work has been developed further by \citep{klimchuk20} where, for 20 loops from the first Hi-C flight data, intensity versus width measurements tended to be uncorrelated or have a direct dependence, implying that the loop flux tube cross sections themselves are approximately circular (assuming that there is non-negligible twist along the flux tube and that the plasma emission is nearly uniform along the magnetic field).

More recently, \citet{williams19} investigate loops from five regions within the FOV of the latest Hi-C flight but at 172\,\AA\ wavelengths \citep[termed Hi-C\,2.1;][]{instpaper}. As with \citet{aschwanden17}, coronal strand  widths of $\approx513$~km were determined for four of the five regions analysed. The final region, which investigates low emission, low density loops, finds much narrower coronal strands of $\approx388$\,km, placing those structures beneath the width of a single AIA pixel. The fact that these strands are above the smallest spatial scale at which \hic\ can resolve individual structures ($220-340$\,km; \citealp{instpaper}) suggests that \hic\ may be beginning to resolve a key spatial scale of coronal loops.

Notably, and the focus for this work, \citet{williams19} also find example structures that may not be fully resolved within the \hic\ data. These relate to smaller `bumps' or turning points in the intensity profiles that are larger than the observational error bars but do not constitute a full, completely isolated strand. Could these be the result of projection effects of overlapping structures along the integrated line of sight for this optically thin plasma, or are they the result of further structures beneath even the resolving abilities of \hic ?

Thus, this current paper outlines approaches to further investigate the possible spatial scale of \hic\ coronal strands reported upon by \citet{williams19} but are not fully resolved as defined above. In \S\ref{sec:data} the \hic\ data preparation is discussed including the Gaussian fit method employed to estimate the width of these partially resolved coronal features. The resulting distribution of fitted widths is described in \S\ref{sec:res} with conclusions reached on the analysis outlined in \S\ref{sec:conc}.
\section{Data Preparation and Analysis Method}\label{sec:data}
On 29\textsuperscript{th} May 2018 at 18:54 UT, \hic\ was successfully relaunched from the White Sands Missile Range, NM, USA, capturing high-resolution data (2k$\times$2k pixels; $4.4^\prime\times4.4^\prime$ field of view) of target active region \ar\ in EUV emission of wavelength 172\,\AA\ (dominated by Fe IX emission $\approx0.8$\,MK) with a plate scale of 0.129\arc. During the flight \hic\ captured 78 images with a 2s exposure time and a 4.4s cadence between 18:56 and 19:02 UT. Full details on the \hic\ instrument can be found in \citet{instpaper}.

\begin{figure*}
\centerline{\includegraphics[width=0.9\textwidth]{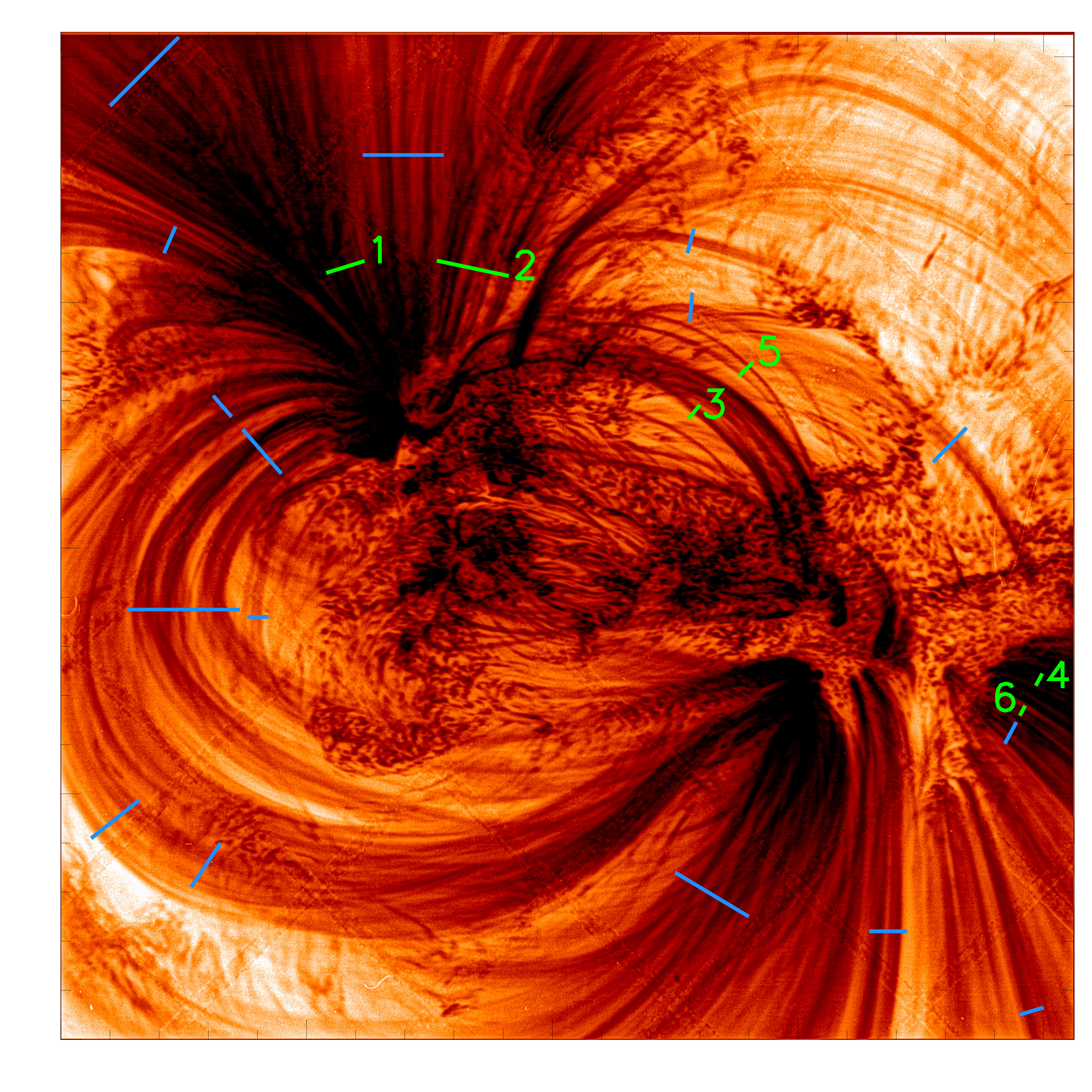}}
\caption{Reverse colour image showing the \hic\ field of view which has been time-averaged for $\approx60$~s and then, for the purpose of this figure only, sharpened with Multi-Scale Gaussian Normalisation \citep[MGN]{morgandruckmuller14}. The locations of the cross section slices where multiple-peaked structures are observed are shown in \textit{blue}, whilst the \textit{green} slices are displayed more clearly in Figure\,\ref{fig:closeup}. The cross-sectional profiles for the slices numbered 1\,-\,4 (5 and 6) are shown in Figure\,\ref{fig:example1} (\ref{fig:fwhmcomp}).}
\label{fig:fov}
\end{figure*}
\begin{figure*}
\centerline{\includegraphics[width=0.9\textwidth]{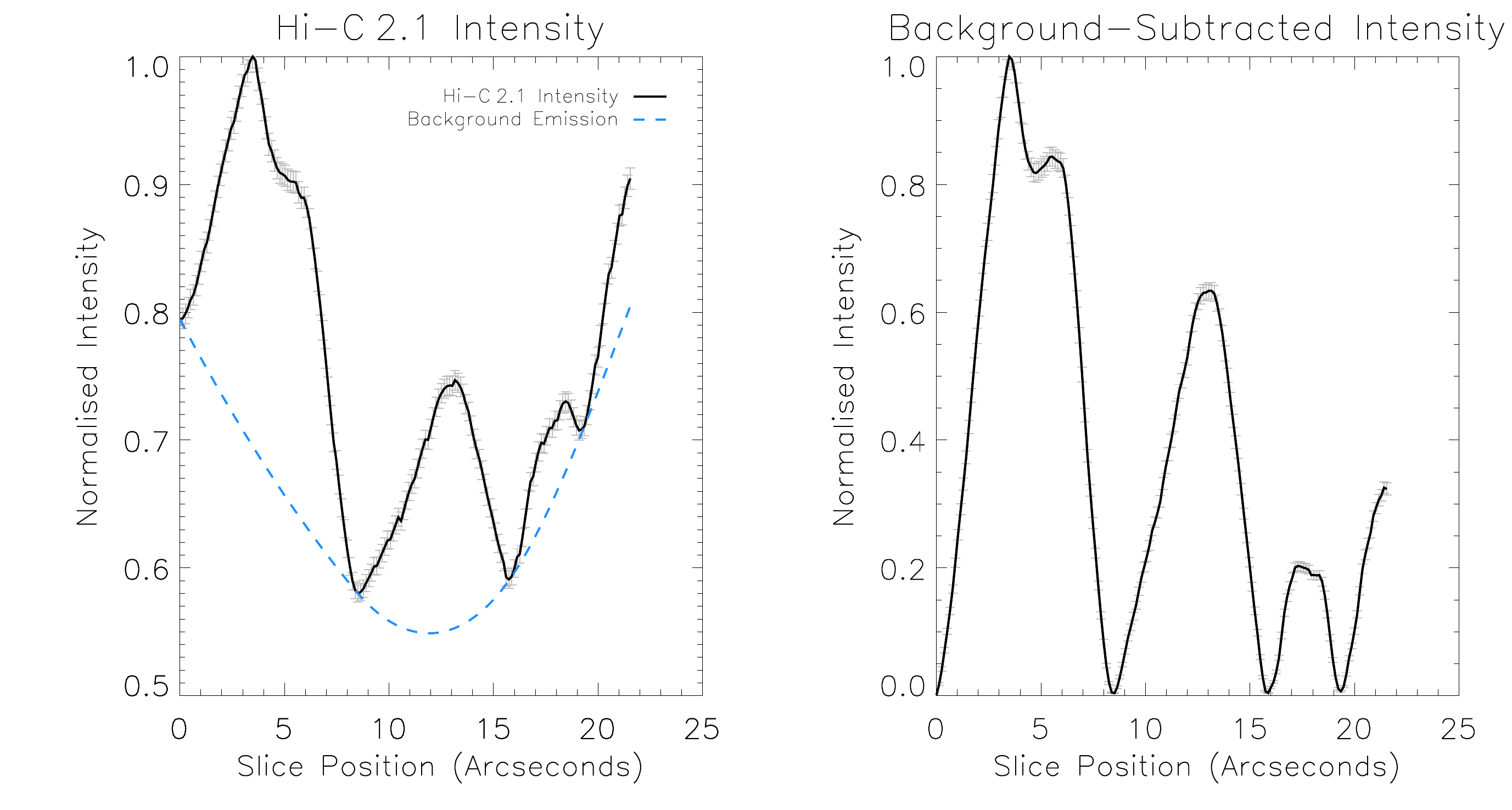}}
\caption{Left: an example cross-section slice from the \hic\ field-of-view (slice\,2 in Figure\,\ref{fig:fov}) is shown in black. The global background obtained by interpolating with a cubic spline through the inflection points is shown via the dashed blue line. Right: the isolated coronal strands that are obtained by subtracting the global background from the \hic\ intensity. The error bars denote five times the Poisson error (\textit{grey}) due to the small magnitude (mean: $1.8\times10^{-3}$) making them difficult to see without magnification.}
\label{fig:background}
\end{figure*}

\subsection{Dataset Extraction and Background Subtraction}\label{sec:bkgd}
The basis of the sample data-set investigated here include a number of subsets from the ten higher-emission cross-section slices analysed by \citet{williams19} plus nine other additional slices from within the \hic\ field of view (see Figure~\ref{fig:fov} where all dataset locations are indicated). In each case the resulting emission profile across the structures would indicate sub-structure strands that are not fully resolved i.e., a non-Gaussian shape.

Following the method outlined in \citep{williams19}, the \hic\ dataset under consideration is time-averaged over a period $\approx60$\,s that is free from spacecraft jitter\footnote{A consequence of the instability experienced during the \hic\ flight is that ghosting of the mesh could not be avoided \citep{instpaper}. This leads to the diamond patterns across the entire \hic\ field-of-view, which are exaggerated when the data is enhanced with MGN (Figure\,\ref{fig:fov}).}. Each cross-section normal to each strand is taken to be 3-pixels deep and the background emission is then subtracted. As outlined in Figure~\ref{fig:background}, this background subtraction is performed by firstly finding all the local minima of a slice, and interpolating through these values using a cubic spline \citep{yi15} to obtain a global trend (dashed blue line). The global trend is then subtracted from the intensity profile along the slice, leaving behind the background subtracted coronal strands (similar to \citealp{aschwanden11,williams19}). Due to the large number of counts detected by \hic\, the Poisson error associated with these isolated coronal strands is minimal (Figure\,\ref{fig:background}).

\subsection{Gaussian Fitting and FWHM Measurements}\label{sec:gaussfit}
The analysis method is based on the assumption that at rest, an isolated coronal strand element has an observed emission profile across its width and normal to the strand axis that is approximately Gaussian. It is important to note that as indicated by \citet{pontin17}, instantaneously coronal strands may not necessarily have a clear Gaussian cross-section. On the other hand, \citep{klimchuk20} have shown from Hi-C observations that coronal strands are likely to have circular cross-sections. To attempt to address this and as indicated previously, the data samples are time-averaged over $\approx60$\,s (the first 11 \hic\ frames) to average out any short timescale changes. Whilst no obvious signs of motion within the structures analysed are noticed in this 60s window, the authors acknowledge that as indicated by \citet{morton13}, small amplitude oscillations could be present which would lead to the measured widths being broader than the structural width due to the time integration performed.

\begin{figure}[!htbp]
\includegraphics[width=0.97\columnwidth]{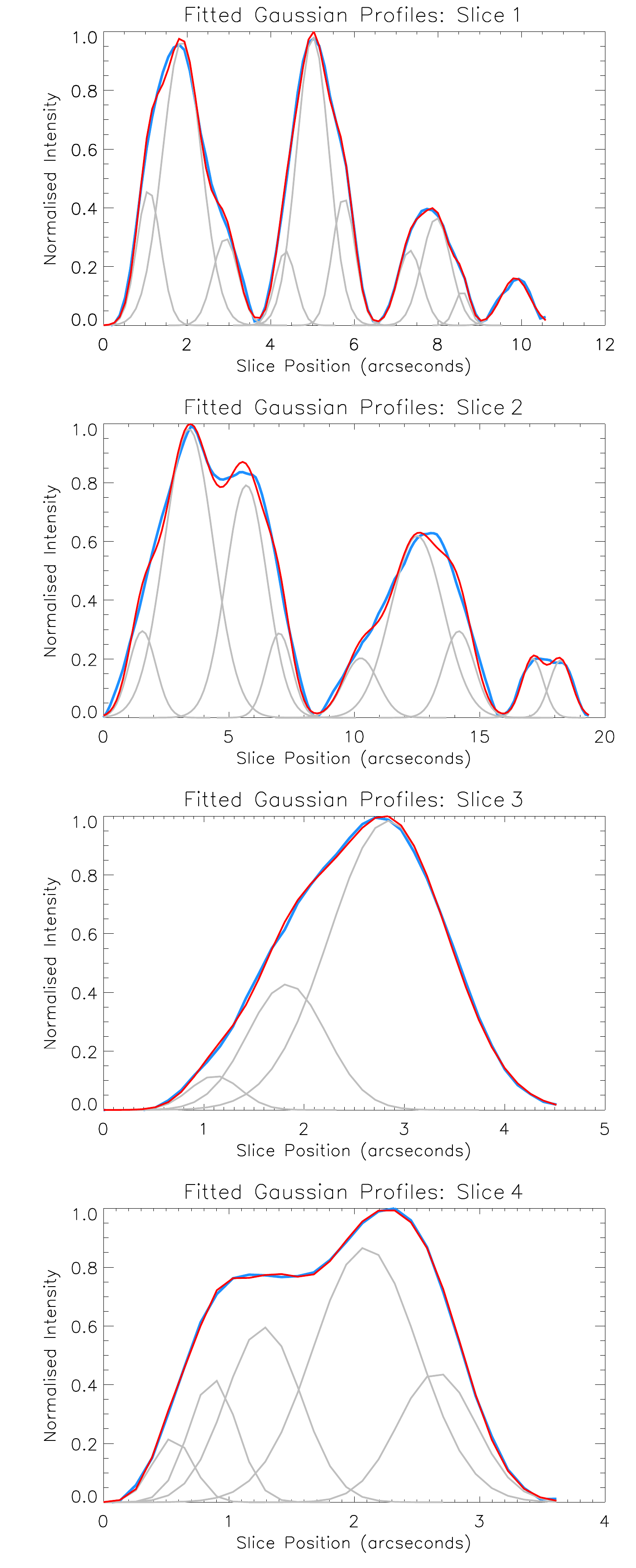}
\caption{The intensity profiles of the \hic\ cross-sectional slices (1\,-\,4 in Figure\,\ref{fig:fov}) are shown in blue. The Gaussian profiles (grey) generated by the non-linear least-squares fitting algorithm and the subsequent fit, $f\left(x\right)$ (red) are also plotted for the four example slices.}
\label{fig:example1}
\end{figure}

Previous width studies (such as \citealp{aschwanden17,williams19}) would have considered the features under examination here (e.g. those between $0\arc-6.5\arc$ in Figure~\ref{fig:example1}) as individual, whole structures in spite of their outline. However, due to their distinct non-Gaussian cross section that is itself well resolved by \hic\,, in this study they are considered to be subsequently modelled as the combination of several Gaussian-shaped coronal strands.

The observed \hic\ intensity profile of a cross-sectional slice is reproduced by simultaneously fitting Gaussian profiles, the number of which is determined by the Akaike Information Criterion \citep[AIC]{akaike74} along with a corrective term (AICc) for small sample sizes. This is fully described in Appendix \ref{app:aic}. Subsequently, the full width at half maximum (FWHM) of the Gaussian profile is measured to provide an estimate of the possible width of the sub-structures likely present within the \hic\ data.

Thus, the method employed to fit Gaussian profiles to the observed \hic\ intensity is as follows. Firstly, the following expression for a Gaussian function, $Y_G$ is used:
\begin{equation}\label{eq:gaussian}
Y_{G} = A \exp\left(\frac{-(x-x_{p})^2}{2W^2}\right),
\end{equation}
whereby $x$ is position along the cross-section slice, $A$ and $x_{p}$ are the amplitude and location of the peak, and $W$ is the Gaussian RMS width. This can be related to the FWHM by: $FWHM=2\,\sqrt{2\ln{2}}\,W\approx2.35\,W$.

An estimate is made on the number of structures, $N$ that could be present within the intensity profile along with their approximate location, width, and amplitude. Summing the $Y_G$ values for $N$ number of Gaussian curves at each pixel yields the model fit:
\begin{equation}\label{eq:fit}
f\left(x\right) = \sum_{i=1}^N Y_{G\left(i\right)}\left(x\right).
\end{equation}

The closeness of the fit at each pixel, $\chi^2\left(x\right)$ is then determined by measuring the deviation of the fit from the original intensity:
\begin{equation}\label{eq:acc}
\chi^2\left(x\right) = \left(\frac{f\left(x\right)-y\left(x\right)}{\sigma\left(x\right)}\right)^2,
\end{equation}
where $y\left(x\right)$ and $\sigma\left(x\right)$ are the observed \hic\ intensity and Poisson error at each pixel. The overall closeness of fit is then taken as $\sum \chi^2\left(x\right)$, which is then reduced to its smallest value by simultaneously adjusting the free parameters $A$, $x_p$, and $W$ for the $N$ Gaussian curves in $f\left(x\right)$. The minimisation of $\sum \chi^2\left(x\right)$ is performed by using the non-linear least-squares curve fitting method, \textit{MPFIT}\footnote{\textit{MPFIT} is freely-available at: http://purl.com/net/mpfit} \citep{mpfit} which is based on the MINPACK-1 FORTRAN library \citep{minpack}. During the fitting process the one-$\sigma$ uncertainties are returned from \textit{MPFIT}. These error values are only accurate if the shape of the likelihood surface is well approximated by a parabolic function. Whether fitting multiple Gaussian profiles to each slice satisfies this condition or not would require analysis beyond the scope of this study, however, it is likely the one-$\sigma$ uncertainties do provide a lower-bound of the FWHM errors.

To determine the appropriate number of Gaussian profiles, $N$ within a given slice, the AIC model selection is employed. This is done by firstly generating several candidate models, where the number of Gaussian curves differs in each model. The non-linear least-squares curve fitting method is then employed for each candidate model and finally the AICc is then computed. The model with the smallest AICc value is then selected as the preferred model for that \hic\ slice.

Once the number of Gaussian profiles contained within a \hic\ slice is determined, the strand width(s) are taken as the Gaussian FWHM value(s). As with previous loop-width studies \citep{aschwanden17,brooks13,brooks16,peter13,williams19} the width measurements are then collated into statistical samples in order to deduce if key structural widths can be extracted from the data.

\section{Results \& Analysis}\label{sec:res}
\begin{figure*}
\centerline{\includegraphics[width=0.95\textwidth]{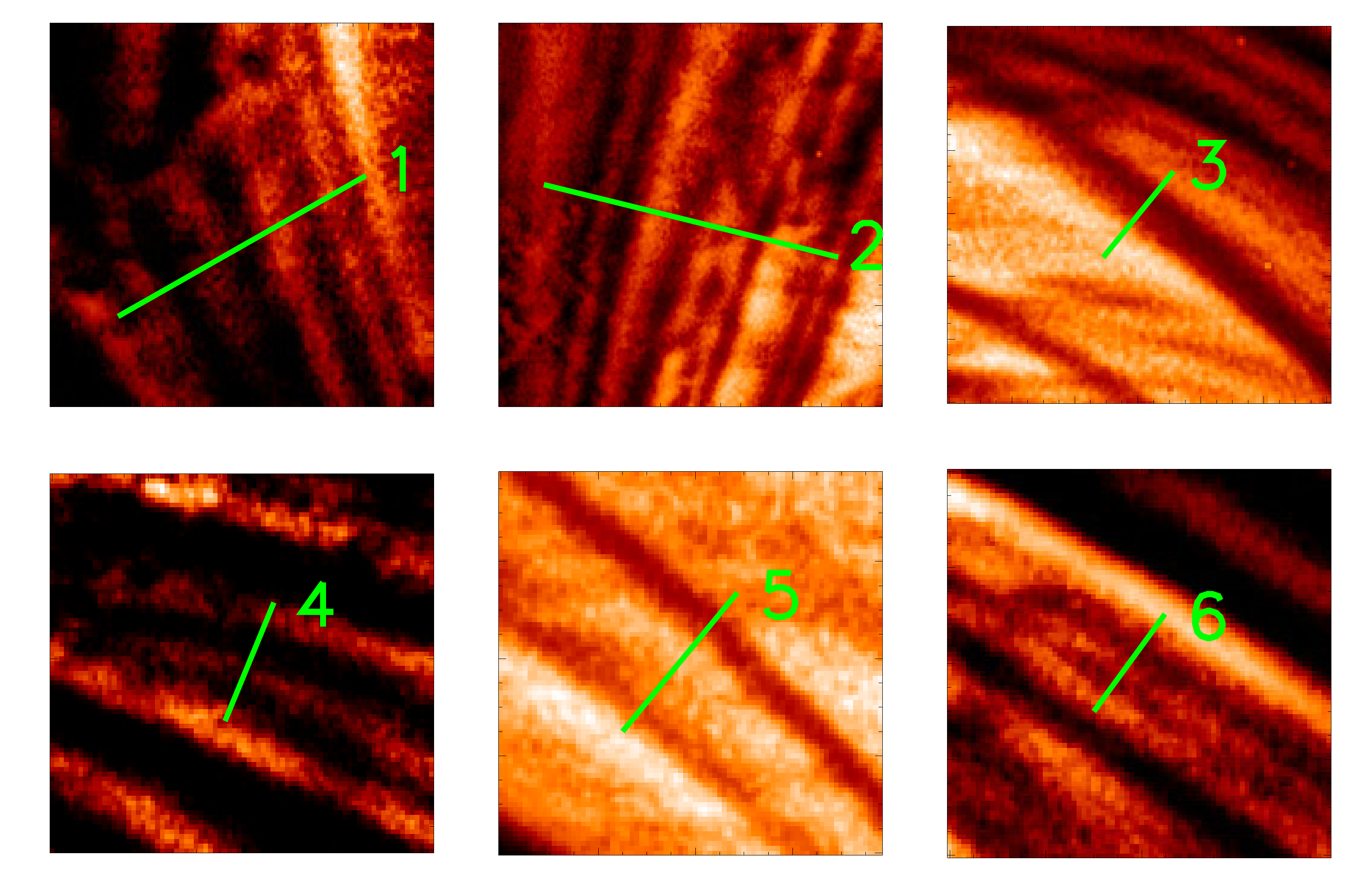}}
\caption{A close-up view of slices numbered 1 - 6 in Figure\,\ref{fig:fov}, which have been sharpened using MGN. The colour tables are normalised to each sub-region and are shown in reverse colour. The cross-sectional profiles of these slices are shown in Figures\,\ref{fig:fwhmcomp} and \ref{fig:example1}.}
\label{fig:closeup}
\end{figure*}
\begin{figure*}
\centerline{\includegraphics[width=0.95\textwidth]{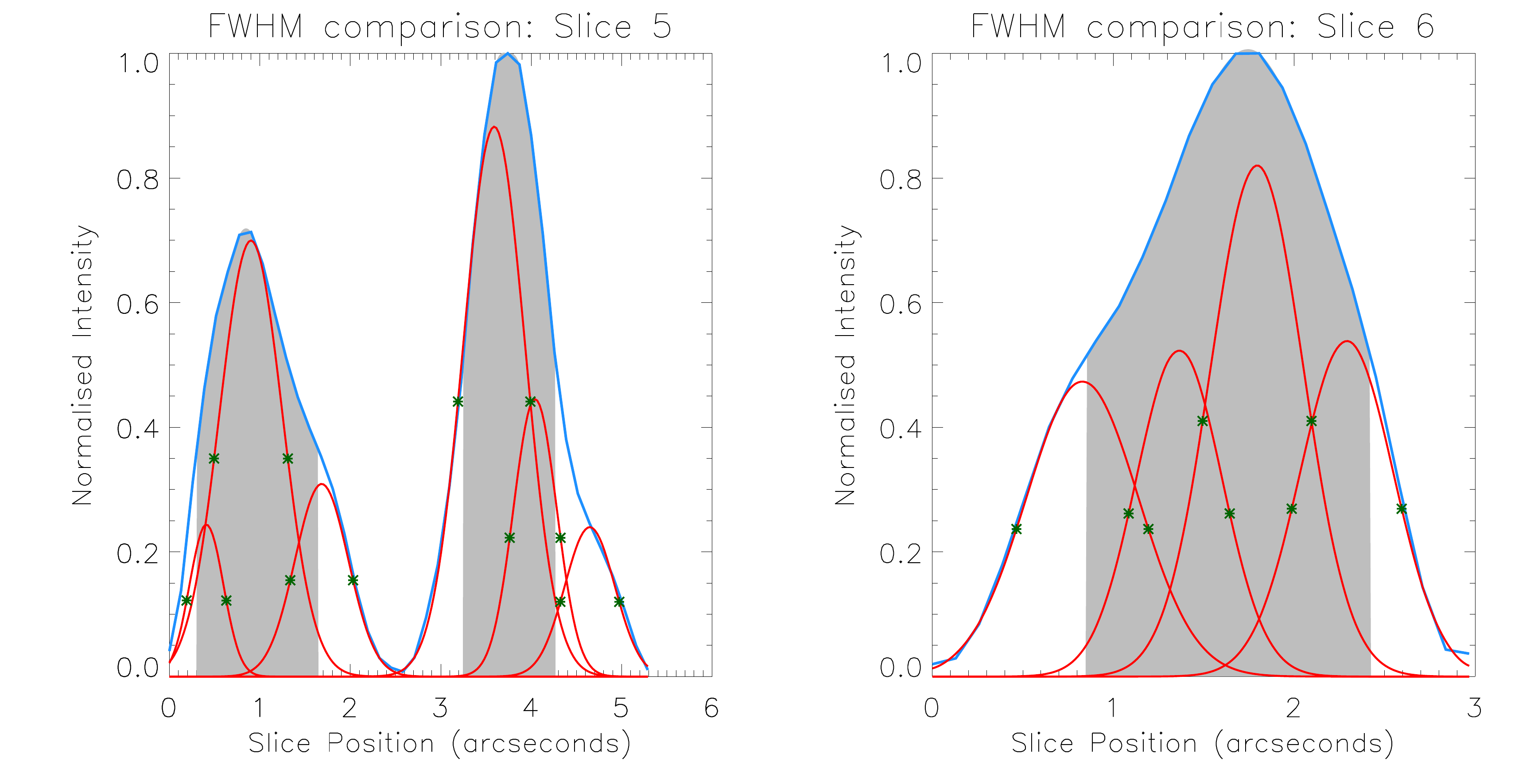}}
\caption{A comparison between the FWHM measurement method of this paper with that of \citet{williams19} for slices 5 and 6 shown in Figure\,\ref{fig:fov}. The \textit{blue} line is the background-subtracted \hic\ intensity, the Gaussian profiles obtained from using the non-linear least-squares fitting method are shown in \textit{red} and their FWHM are denoted by the \textit{green} asterisks. The solid \textit{grey} bands indicate the FWHM of the \hic\ structures as determined by our previous method \citep{williams19}.}
\label{fig:fwhmcomp}
\end{figure*}
\begin{figure}
\includegraphics[width=\columnwidth]{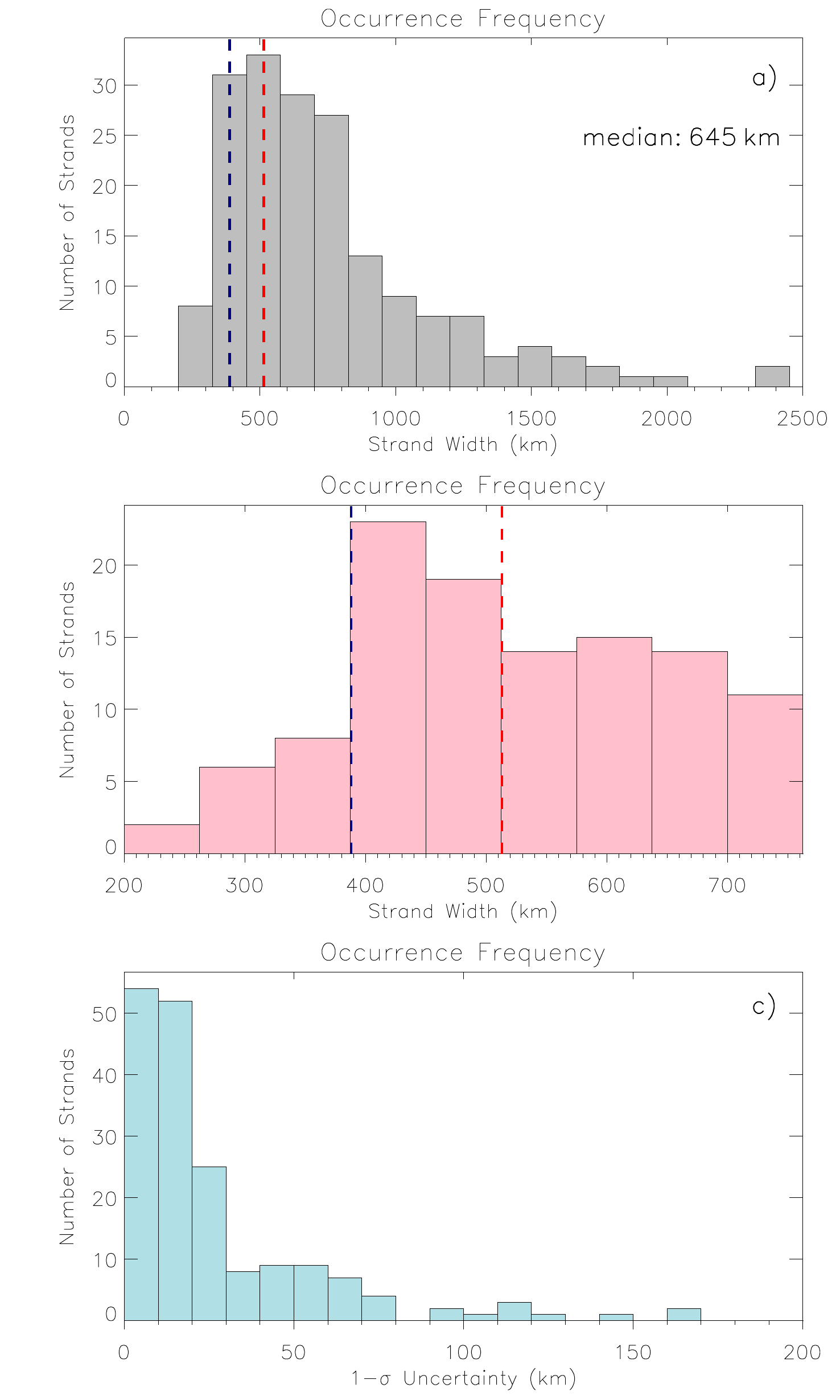}
\caption{a) Occurrence frequency plot of the 183 FWHM measurements for the fitted Gaussian curves with a bin width of 125\,km ; b) a subset of the Gaussian profiles plotted at bin widths of 62.5\,km that correspond to the most populous widths in a). The dashed vertical navy (red) lines indicate the low (high) emission strand widths obtained in \citet{williams19}. Panel c) shows the $1-\sigma$ errors for the 183 widths obtained for the Gaussian profiles in a).}
\label{fig:fwhm}
\end{figure}

Employing the \cf\ discussed in the previous section, a total of 183 Gaussian profiles are fitted to twenty four \hic\ cross-sectional slices. As seen in the field-of-view plot (Figure\,\ref{fig:fov}) it is not easy to completely isolate a coronal strand. For example, to the north of slice 5 (Figure\,\ref{fig:closeup}) there is an increase in intensity due to a crossing of another emitting feature along the integrated line-of-sight. Care is taken to avoid contamination from such structures, though it is possible that some residual emission may remain in the slices selected. However, the relative intensity of the much brighter strand to the often weaker contaminating emission means that it is removed during background subtraction.

The following subsections compare the twenty four cross-sectional slices to those outlined in \citet{williams19} as well as examining the frequency distribution of the newly fitted Gaussian profiles.

\subsection{FWHM Method Comparison}\label{sec:paper1}
Here, comparison is made between the resulting strand widths obtained by fitting multiple Gaussian profiles to \hic\ structures versus the widths obtained using the previous method \citep{williams19} now with the improved background subtraction discussed in \S\,\ref{sec:bkgd}. Two examples are outlined (Figure\,\ref{fig:fwhmcomp}) where a non-Gaussian distribution is seen. For slice 5 (6), the AICc determined that six (four) Gaussian profiles are supported by the \hic\ data whereas measuring the widths of the non-Gaussian profiles provides two (one) structures.

The \citet{williams19} method provides widths of $\approx745$\,km and $980$\,km yielding a mean of $\approx860$\,km for slice 5. Comparatively, the \cf\ yields minimum and maximum widths of $\approx320$\,km and $\approx590$\,km, and a mean of $\approx480$\,km. The width of the single structure in slice 6 measured with the \citet{williams19} method is $\approx1145$\,km. As with slice 5, the \cf\ provides narrower widths with the minimum, maximum, and mean now being $\approx405$\,km, $\approx530$\,km, and $\approx450$\,km, respectively.

The width estimates of the structures centred at 1\arc (slice 5) and 10.4\arc (slice 6) may be artificially broadened by their shape using the method employed by \citet{williams19} due to the observable change in gradient that occurs in the vicinity of the half-maximum intensity value. The structure at 3.8\arc (slice 5) does not appear to be affected by this as the change in gradient (or `bump') occurs much lower along the structure than the half-maximum intensity value. However, the measured width of this structure is still $\approx300$\,km broader than the maximum AICc determined Gaussian width, which indicates previous analysis methods may have over-estimated the strand widths of structures that are potentially not completely isolated from the background and/or other structures along the integrated line-of-sight.

\subsection{Distribution of Fitted Widths}\label{sec:diswid}
 In Figure \ref{fig:example1}, the cross-sectional profiles (\textit{blue}) are shown of the \hic\ slices numbered 1\,-\,4 in Figure\,\ref{fig:fov} along with the best AICc-determined fits and Gaussian profiles generated from the \cf, shown in \textit{red} and \textit{grey} respectively. From the four examples shown here, it is seen that there is good agreement between the observed \hic\ intensity and the generated fit though some minor discrepancies may occasionally occur (e.g. slice 2 between 3.5\arc and 6.5\arc). These discrepancies could be eradicated by adding additional Gaussian profiles along the slices; however, the additional parameters introduced are not supported by the AICc model selection.

In Figure\,\ref{fig:fwhm}\,(a) the FWHM values of the 183 Gaussian profiles are collated into an occurrence frequency plot binned at $125$\,km intervals so as to be consistent with the previous study \citep{williams19} where $1\arc \approx 725$\,km. A sub-section of this data ($200-760$\,km) is shown in Figure\,\ref{fig:fwhm}\,(b) which is binned at half the spatial scale of Figure\,\ref{fig:fwhm}\,(a) ($62.5$\,km). Figure\,\ref{fig:fwhm}\,(c) shows the one-$\sigma$ errors for the 183 Gaussian widths indicating the majority of errors are $\lesssim50$\,km.

The distribution of all analysed widths in Figure\,\ref{fig:fwhm}\,(a) reveals that the most populous widths are between $450-575$\,km; this matches the high-emission region results from \citet{williams19}. The median width for this data is $645$\,km and is akin to that obtained by \citet{brooks13}; however, this value is due largely to the presence of a number of broader strands ($>1000$\,km). Furthermore, $\approx21\%$ of widths exceed $1000$\,km whilst $\approx32\%$ of the strands studied are at the SDO/AIA resolving scale of $600-1000$\,km. From this, $\approx47\%$ of the strands are beneath the resolving scale of AIA.

Figure\,\ref{fig:fwhm}\,(a) reveals the most populous strand widths in this study occur between $\approx200-760$\,km with the number of width samples above this spatial scale rapidly decreasing. Figure\,\ref{fig:fwhm}\,(b) shows a subset of the obtained widths having been re-binned to 62.5\,km intervals, which allows for further insight on the distribution of widths for the most populous occurrence frequency bins of Figure\,\ref{fig:fwhm}\,(a). The obtained \hic\ strand widths reveal the presence of numerous strands ($\approx32\%$ of the 183 Gaussian widths) whose FWHM are beneath the most frequent high-emission strand widths seen previously \citep[$\approx513$\,km]{williams19}. Similarly, $\approx17\%$ reside beneath an AIA pixel width of 435\,km. Comparatively then, only $\approx6\%$ of the strands are actually at the the scale at which \hic\ can resolve structures \citep[$\approx220-340$\,km]{instpaper}, indicating that current instrumentation may now be beginning to observe a prevalent spatial scale.

As with \citet{williams19}, this analysis reveals that the most-likely strand widths of $450-575$\,km are typically of the order of an AIA pixel width. This result coupled with the low-percentage ($\approx6\%$) of strands at the \hic\ resolving scale suggests that the non-Gaussian structures observed are predominately the result of multiple, potentially resolvable strands overlapping along the integrated line-of-sight rather than the result of finer strands that even \hic\ is unable to resolve into distinct features. It should be noted that these results are $50-250$\,km narrower than previous Hi-C findings that focused on 193\,\AA\ emission \citep{aschwanden17,brooks13}.

Nevertheless, this does not rule out the possibility of strands within or beneath the resolving power of \hic. For example, using CRISP H-$\alpha$ data, \citet{scullion14} find that the most populous strand width is $\approx100$\,km. However, the temperature of those structures is 1-2 orders of magnitude lower than that observed with \hic.

\section{Summary and Conclusions}\label{sec:conc}
This work outlines a follow-up analysis to \citet{williams19} where non-Gaussian shaped width profiles that are not fully resolved within the \hic\ data are further investigated. To estimate the widths of possible strands, Gaussian functions are first fitted to approximate the \hic\ intensity profiles using the method outlined in \S\ref{sec:gaussfit}. The non-linear least-squares curve fitting method employed automatically determines the Gaussian RMS width due to $W$ being a free parameter (Equation~\ref{eq:gaussian}) used to reduce $\sum \chi^2\left(x\right)$. The number of Gaussian profiles and subsequently the number of RMS widths measured are determined by the AICc, which are then converted to FWHM for our width analysis study.

The FWHM are collated into occurrence frequency plots (Figure~\ref{fig:fwhm}) revealing the most frequent strand width is $\approx450-575$\,km. The spatial scales obtained in this study largely agree with previous findings \citep{aschwanden17,williams19} where typical widths the size of an AIA pixel are seen. Additionally, the results reveal that only $\approx6\%$ of the strands analysed reside at the smallest spatial scales that \hic\ can resolve into distinct structures \citep[$220-340$\,km]{instpaper}. Together, these findings strongly suggest that structures emitting at 172\,\AA\ that cannot be resolved into distinct features by \hic\ are likely to comprise of multiple strands overlapping along the integrated line-of-sight rather than being an amalgamation of strands at/below the resolving scale of \hic. For coronal loop modelling, the onus must now be on the determination of the spatial scale at which heating occurs that leads to the formation of individual magnetic strands that i) have widths $450-575$\,km and ii) are filled with plasma around 1\,MK.

Furthermore, recent work by \citet{klimchuk20} investigated the widths along the length of isolated coronal structures and found no correlation between width and intensity. However, as is noted in \citet{klimchuk20}, if a structure indicated signs of any possible substructure, then that particular example was not included in the study data. Thus, employing the methods adopted in this work on those rejected examples would allow for that type of analysis to be performed along the observable length of coronal structural sub-elements and not only the aforementioned monolithic features. This will be addressed in a follow-up study using the \hic\ data set examined here.

The highly anticipated ESA mission Solar Orbiter (SolO) will provide close-up ($\approx0.28$\,AU), high-latitude (34$^{\circ}$) solar observations. During the mission there will be several observation windows where the spatial resolution of EUV Imager (EUI) HRI as well as the selected passband (174\,\AA) will be similar to that of Hi-C. However, it is likely SolO will have longer observation windows over which any target active region may be studied (Hi-C only captures 2.5 minutes of usable data per flight). This will allow for significantly improved strand width determination across many differing coronal structures.

\begin{appendix}
\section{Akaike Information Criterion}\label{app:aic}

To determine the number of strands that may be hidden within the \hic\ data the Akaike Information Criterion \citep{akaike74} is employed to determine the optimal number of Gaussian profiles supported by the data. Each additional Gaussian that is added to the model introduces an additional three parameters that may be tweaked to better allow for the observed data to be replicated by equations \ref{eq:gaussian}\,-\,\ref{eq:acc}. Employing a model selection method such as AIC helps minimise the possibility of selecting a model with too many(few) Gaussian curves, and thus the danger of over(under)-fitting the \hic\ data.

Often, the AIC is defined as $AIC = 2k -2\ln{\left(L_{max}\right)}$ where $k$ is the number of parameters in the model and $L_{max}$ is the maximum likelihood. In this study, a least-squares model fitting is employed and thus the maximum likelihood estimate for the variance of a model's distribution of residuals is $\hat{\sigma}^2 = RSS/n$, where $n$ is the sample size and $RSS$ is the residual sum of squares:
\begin{equation}\label{eq:rss}
RSS = \Sigma_{i=1}^{n}\left(y_i-f\left(x_i\right)\right)^2.
\end{equation}
Thus, the maximum value of a model's likelihood function can be expressed as:
\begin{equation}\label{eq:llh}
-\frac{n}{2}\ln{\left(2\pi\right)}-\frac{n}{2}\ln{\left(\hat{\sigma}^2\right)}-\frac{1}{2\hat{\sigma}^2}RSS = -\frac{n}{2}\ln{\left(\frac{RSS}{n}\right)}+C,
\end{equation}
where $C$ is an independent constant that does not change unless $y$ does. Following \citet[p.63]{burnham02}, this means the AIC for a least-squares model can be expressed as:
\begin{equation}
AIC = 2k + n\ln{\left(\frac{RSS}{n}\right)}-2C = 2k + n\ln{\left(RSS\right)}-\left(n\ln{\left(n\right)}+2C\right),
\end{equation}
which can be further simplified to
\begin{equation}\label{eq:AICRSS}
AIC = 2k + n\ln{\left(RSS\right)},
\end{equation}
as $\left(n\ln{\left(n\right)}+2C\right)$ is a constant (provided $y$ does not change) and only the differences in AIC are meaningful.

If $n$ is small, AIC may prefer models which have more parameters and lead to over-fitting of the data. As such, a correction for this is to use the AICc, which provides an additional term accounting for $n$ and $k$:
\begin{equation}
AICc = AIC + \frac{2k^2 + 2k}{n-k-1},
\end{equation}
that converges to 0 as $n\xrightarrow{}\infty$ meaning $AICc \equiv AIC$ for large values of $n$.

\subsection{AIC Test Cases}
\begin{figure*}[ht!]
\centering
\begin{minipage}[b]{.6\textwidth}
\centerline{\includegraphics[width=\textwidth]{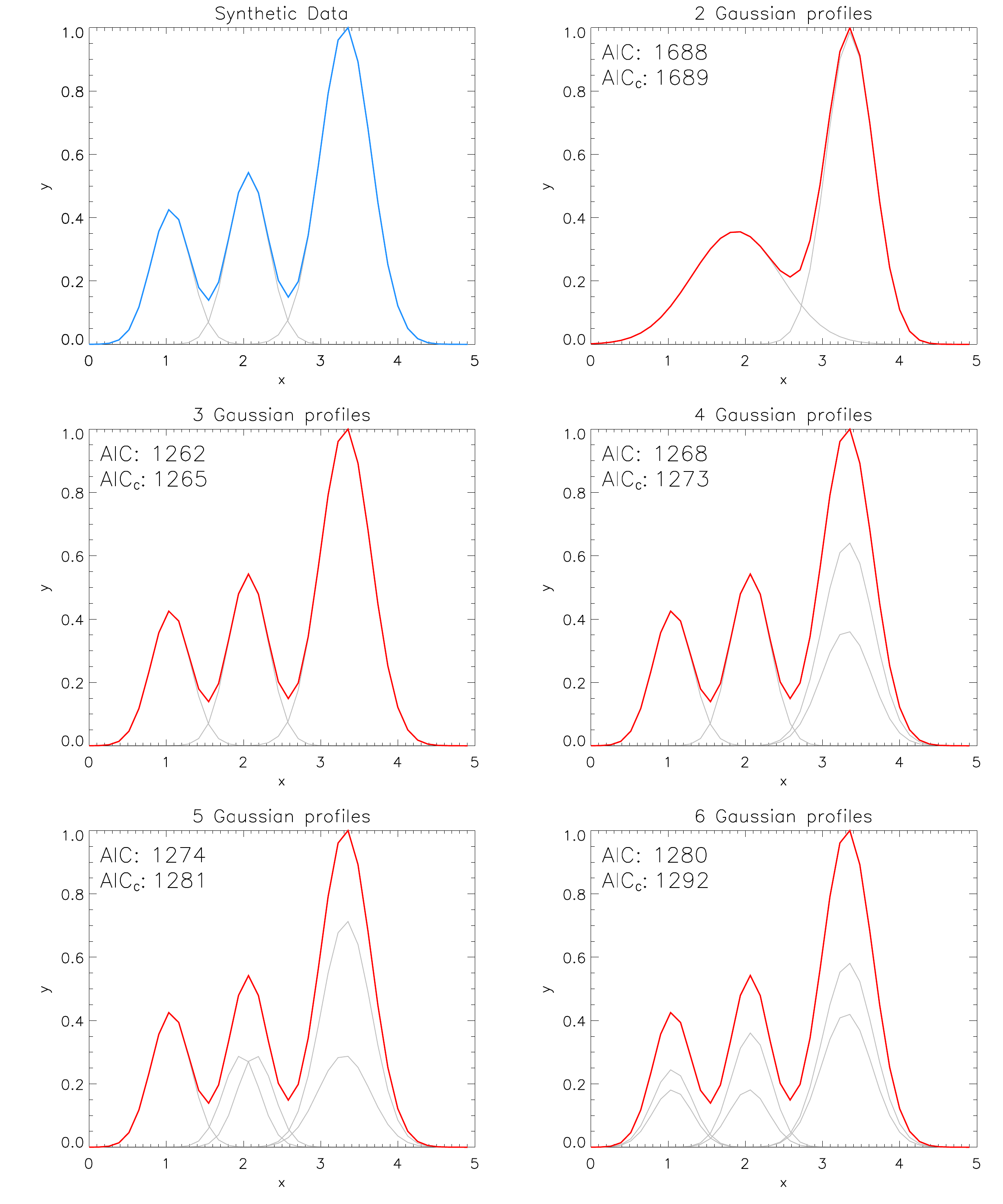}}
\end{minipage}
\caption{The first AIC model selection  test case. The upper-left panel labelled synthetic data is generated using three Gaussian functions. Using the Gaussian fitting method employed in this paper, a number of Gaussian functions (shown in gray) are used to replicate the synthetic data and their AIC and AICc values are indicated.}
\label{fig:test1}
\end{figure*}
\begin{figure*}[ht!]
\centering
\begin{minipage}[b]{.47\textwidth}
\centerline{\includegraphics[width=\textwidth]{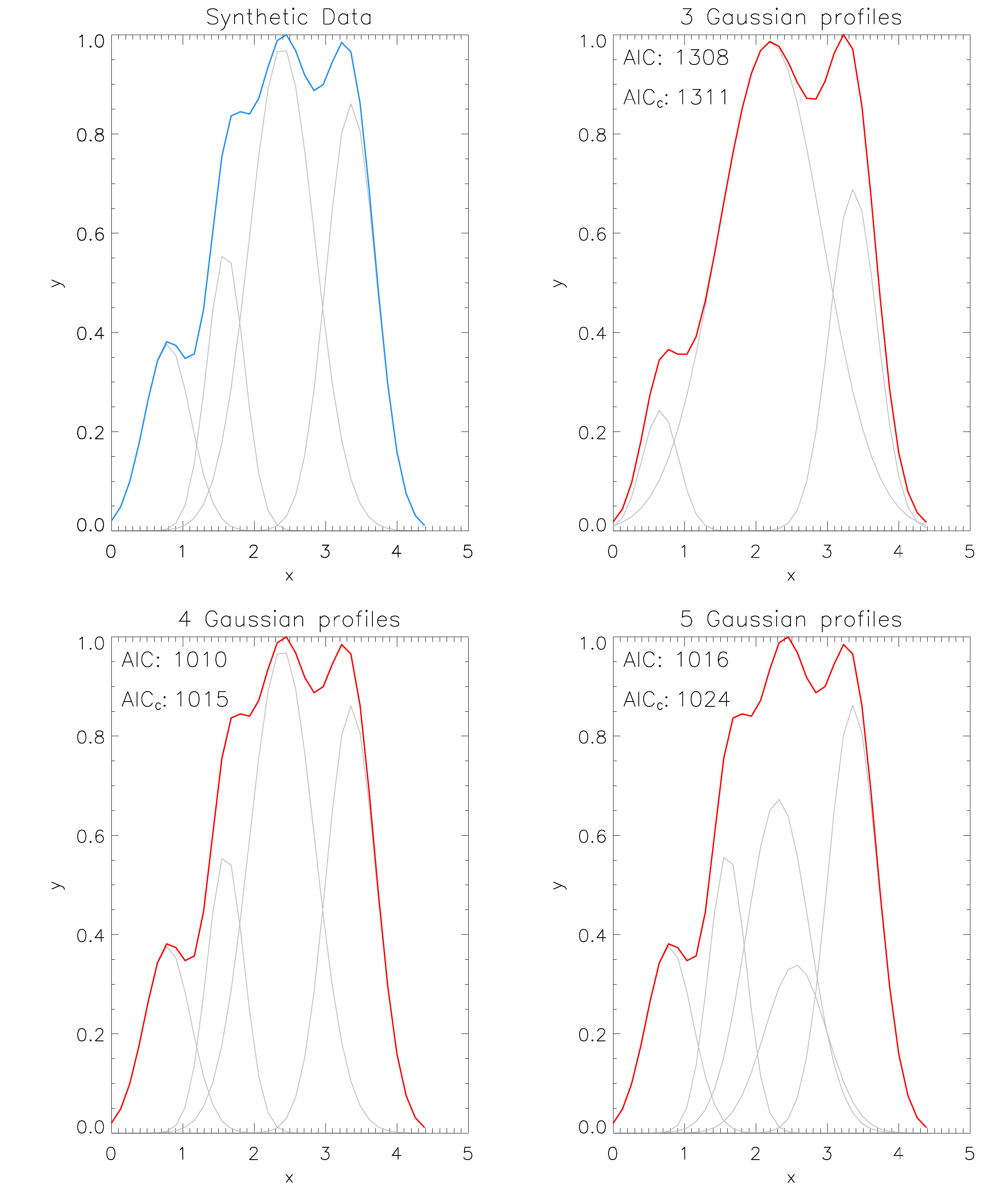}}
\caption{The second AIC model selection  test case. The upper-left panel labelled synthetic data is generated using four Gaussian functions. Using the Gaussian fitting method employed in this paper, a number of Gaussian functions (shown in gray) are used to replicate the synthetic data and their AIC and AICc values are indicated.}
\label{fig:test2}
\end{minipage}
\begin{minipage}[b]{.47\textwidth}
\centerline{\includegraphics[width=\textwidth]{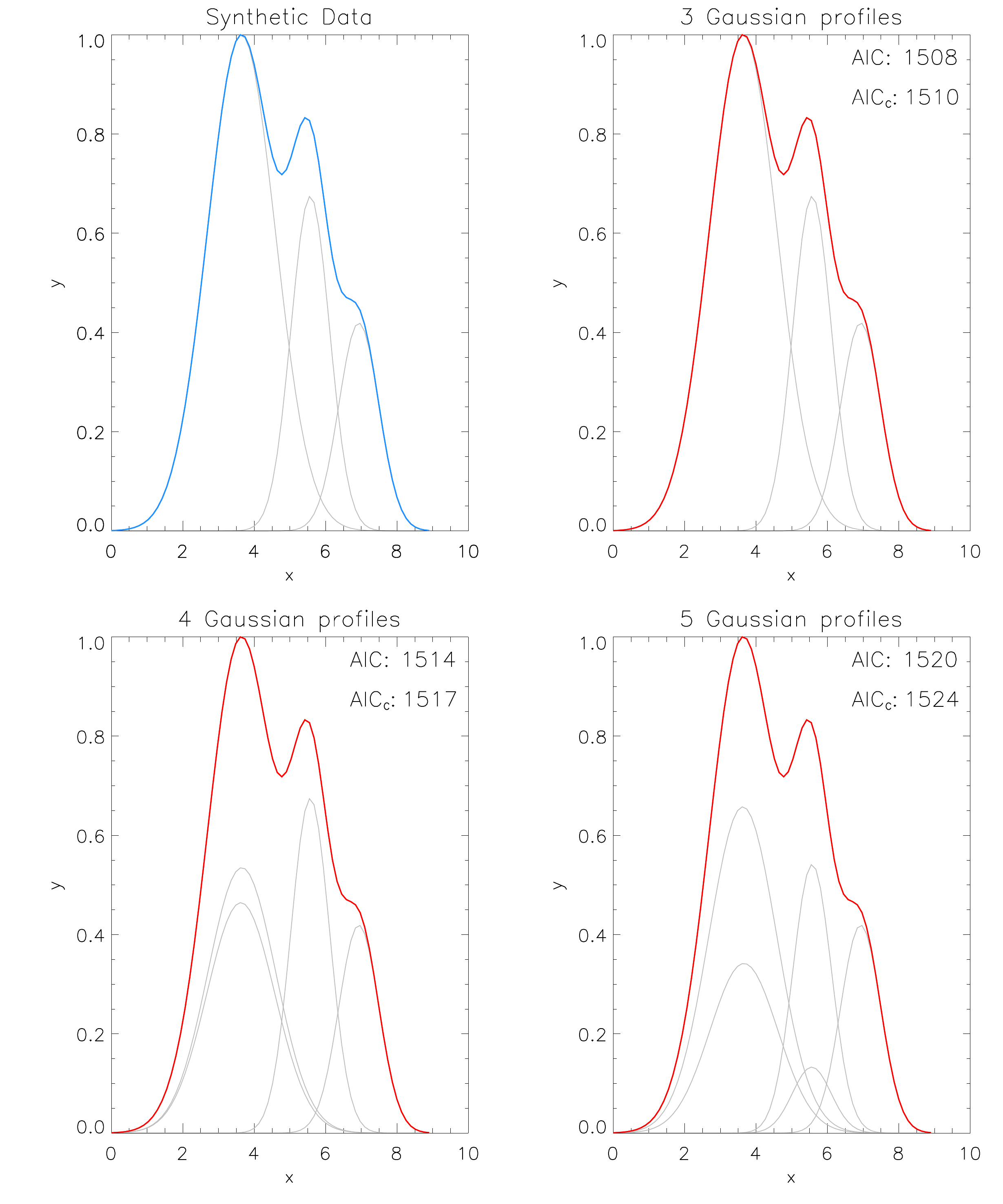}}
\caption{The thrid AIC model selection  test case. The upper-left panel labelled synthetic data is generated using three Gaussian functions. Using the Gaussian fitting method employed in this paper, a number of Gaussian functions (shown in gray) are used to replicate the synthetic data and their AIC and AICc values are indicated.}
\label{fig:test3}
\end{minipage}
\end{figure*}

To validate the AICc model selection, three test cases are devised that are similar to what a \hic\ cross-sectional slice may look like in this study. The test cases are generated by specifying a number of Gaussian profiles using equation \ref{eq:gaussian}, which are then combined using equation \ref{eq:fit} to generate the synthetic data for each test case. This process allows for the AICc model selection accuracy to  be verified as the number of Gaussian profiles required to generate the three test cases are known.

The first test case is composed of three distinct Gaussian profiles which is shown by the blue plot in Figure\,\ref{fig:test1}. An initial guess on the number of Gaussian profiles (2, 3, 4, 5, and 6 Gaussian curves) and their associated free parameters ($A$, $x_p$, and $W$) are made, which are then passed through our non-linear least squares fitting method. During this fitting method, the AIC and AICc values are computed for each model shown in Figure\,\ref{fig:test1}. This reveals that the smallest AIC/AICc values are for the model consisting of three Gaussian functions, which matches the number used to generate the synthetic data. Test cases 2 and 3 (Figures\,\ref{fig:test2}\,and\,\ref{fig:test3}) are more complex than the first test case, and subsequently provide a closer representation of \hic\ data. Again, the lowest AIC and AICc values correspond to the models consisting of four and three Gaussian profiles, which match the number of Gaussian profiles used in generating the test data.

Whilst both the AIC and AICc show agreement on the model selection for the three test cases analysed in this appendix, some of the \hic\ cross-sectional slices selected may be of sufficiently small sample size that AIC would favour over-fitting the data, which would lead to artificially narrow widths of the strands in question. Therefore and as outlined above, the AICc is employed in this case for determining the number of strands within a given cross-sectional slice.
\end{appendix}


\end{document}